\documentclass[12pt]{article}
\usepackage{pic04}
\usepackage{hyperref}
\usepackage{url}
\usepackage{graphicx}

\begin{document}

\title{\bf Direct and Indirect Searches for Dark Matter in the Form of Weakly Interacting Massive Particles (WIMPs)}
\author{
Nader Mirabolfathi        \\
{\em University of California, Berkeley}}
\maketitle

%
%
%
%
%
%
\vspace{4.5cm}
%

\baselineskip=14.5pt
\begin{abstract}
Numerous lines of evidence indicate that the matter content of the Universe
is dominated by some unseen component.  Determining the nature of this Dark
Matter is one of the most important problems in cosmology.  Weakly
Interacting Massive Particles (WIMPs) are widely considered to be one of the
best candidates which may comprise the Dark Matter.  A brief overview of the
different methods being used to search for WIMP Dark Matter is given,
focusing on the technologies of several benchmark experiments.

\end{abstract}
\newpage

\baselineskip=17pt

\section{Introduction}

The nature of dark matter is one of the oldest and most important open 
questions in cosmology, dating back to the first observations of anomalous high 
kinetic energies in
distant galaxy clusters made by Swiss cosmologist Fritz Zwicky in 1933 
\cite{zwicky}. Since then, exciting developments in observations of the cosmic microwave 
background (CMB),
large-scale structure (LSS), and type Ia supernovae have allowed more accurate
measurement of the various parameters of the ``standard model of cosmology''.  In
particular, the Universe appears to be dominated ($\sim70\%$) by the vacuum 
energy density, or Dark Energy, while the remainder ($\sim30\% $) is made up of matter. 
Furthermore, most of the matter of the Universe appears to be non-luminous (i.e. 
dark) and non-baryonic. The nature of the Dark Matter is still unknown, and remains the 
subject of intense research activity.
Weakly Interacting Massive Particles (WIMPs), a generic name for 
heavy particles interacting at the weak scale with baryonic matter, are among the 
best candidates which may comprise the Dark Matter. If such particles are produced 
thermally in the early universe, their weak-scale couplings explains why their relic density 
is of the order of critical density today.  Independently, supersymmetry theories 
predict a stable s-particle state whose properties are very similar to the hypothetical WIMPs.

   This paper is a brief review and update of different experiments 
aiming to detect Dark Matter in the form of WIMPs.  The experiments described in this paper 
are not meant to be a complete list, but are selected to represent broader classes of 
the similar experiments.

\section{Direct WIMP Searches}

WIMPs can be detected either via (very rare) elastic scattering off 
the nucleus of ordinary matter (``direct detection'') or by measuring their annihilation 
products (``indirect detection''). The latter method is the subject of the section 3. In 
this section, the expected spectrum and rate of interaction of WIMPs with ordinary 
matter are estimated. After the experimental challenges for directly detecting WIMPs have 
been introduced, a brief review of the status and results from few important direct 
detection experiments will be presented.

  WIMPs are expected to interact with the nucleons in ordinary matter \cite{goodman}. The WIMP-nucleon 
elastic-scattering cross-section $\sigma_{\mathrm{WIMP-p}}$ is SUSY model 
dependent \cite{jungman}. One of the goals of all WIMP direct detection experiments is to determine 
or to limit $\sigma_{\mathrm{WIMP-p}}$, and thus constrain the free parameter space available for SUSY models.
If the WIMPs are bound to the galaxy by the gravitational force, we 
can assume that their distribution should follow (Boltzman):

\begin{equation}
f(\vec r , \vec v, \vec v_E)=e^{-\frac{M_W (\vec v_E+\vec v)^2+M_W \Phi (\vec r)}{k_B T}}
\label{equ1}
\end{equation}

where $M_W$ and $T$ are the WIMP mass and the equivalent temperature ($kT=1/2 M_W v_{0}^{2}$), $\Phi(\vec r)$ is the local gravitational potential, $v$ is WIMP velocity with respect to the earth, and  $v_E$ is the velocity of the Earth with respect to the center of the galaxy (sum of the Sun's velocity with respect to the center of the galaxy and the 
Earth's velocity with respect to the Sun):

\begin{equation}
v_E = 232 + 15 \cos{2\pi \frac{t-152.5_{\mathrm{days}}}{365.25_{\mathrm{days}}}} \;\;\;[\textrm{km/s}]
\label{equ2}
\end{equation}

Since the sinusoidal behavior comes from the Earth's motion around 
the sun, we expect its amplitude to be determined by the relationship between the Sun's 
velocity vector and the plane of the Earth's orbit.  Thus we should expect an annual 
modulation of the WIMP flux of $\sim \pm  6\%$.
The event rate is the product of the number of target nuclei ($mN_0/A$), 
the incoming flux of WIMPs, $v\cdot n$ (n determined by the eq. \ref{equ1} from cosmology),  and 
$\sigma_{\mathrm{WIMP-p}}$ (predicted from the SUSY model in play). One could integrate over the WIMP 
velocity distribution and obtain the overall interaction rate. However, we are more 
interested in deriving the differential recoil energy spectrum. We will see that such a spectrum 
will directly give $\sigma_{\mathrm{WIMP-p}}$  and the mass of the WIMPs. The recoil energy of a nucleus of 
mass $M_T$, which is hit by a WIMP of energy $E$, and which is recoiling at an angle $\theta$, is given by:

\begin{equation}
E_R = E \frac{4 M_W M_T}{(M_W + M_T)^2} \frac{(1-\cos{\theta})}{2}
\label{equ3}
\end{equation}

Assuming a hard sphere scattering model (uniform $E_R$ distribution) we 
can calculate the differential rate as:

\begin{equation}
\frac{\partial{R}}{\partial{E_R}} = \frac{R_0}{rE_0} \frac{2\pi^{3/2}v_0}{K} \int_{v_{min}}^{v_{esc}} v  e^{\frac{(v+v_E)^2}{v_{0}^{2}}}dv
\label{equ4}
\end{equation}

\begin{equation}
r = \frac{4M_WM_T}{(M_W+M_T)^2}
\label{equ4b}
\end{equation}

where $R_0 \sim n_0v_0 \sigma_{\mathrm{WIMP-p}}$ , $E_0=k_BT$,  $K$ is a normalization factor, $v_{min}$ is the minimum WIMPs 
velocity necessary to produce a recoil of 
$E_R$ and $v_{esc}$ is the galactic escape velocity. At the limit conditions ($v_e=0$ and $v_{esc} =\infty$), eq. \ref{equ4} becomes:

\begin{equation}
\frac{\partial{R}}{\partial{E_R}} = \frac{R_0}{rE_0} e^{-\frac{E_R}{rE_0}}
\label{equ5}
\end{equation}

 This form, although incorrect, illustrates the typical exponential 
behavior of the recoil energy spectrum. In particular, it shows that by observing the 
amplitude and the shape of
the recoil energy spectrum, it is possible to constrain two 
physically important parameters: $M_W$ and $\sigma_{\mathrm{WIMP-p}}$.

\begin{figure}[htbp]
  \centerline{\hbox{ \hspace{0.2cm}
    \includegraphics[width=6.5cm]{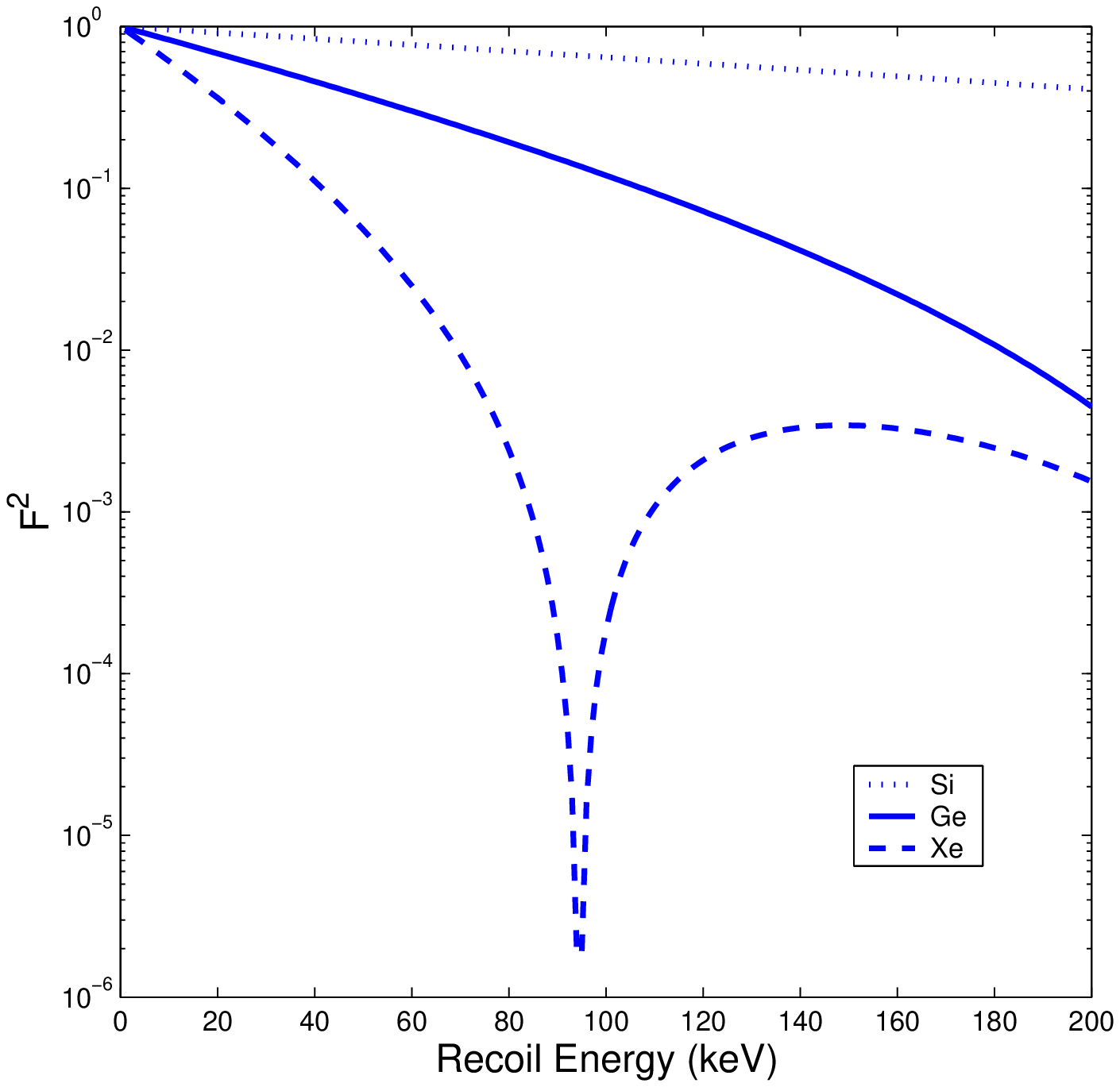}
    \hspace{0.3cm}
    \includegraphics[width=6.5cm]{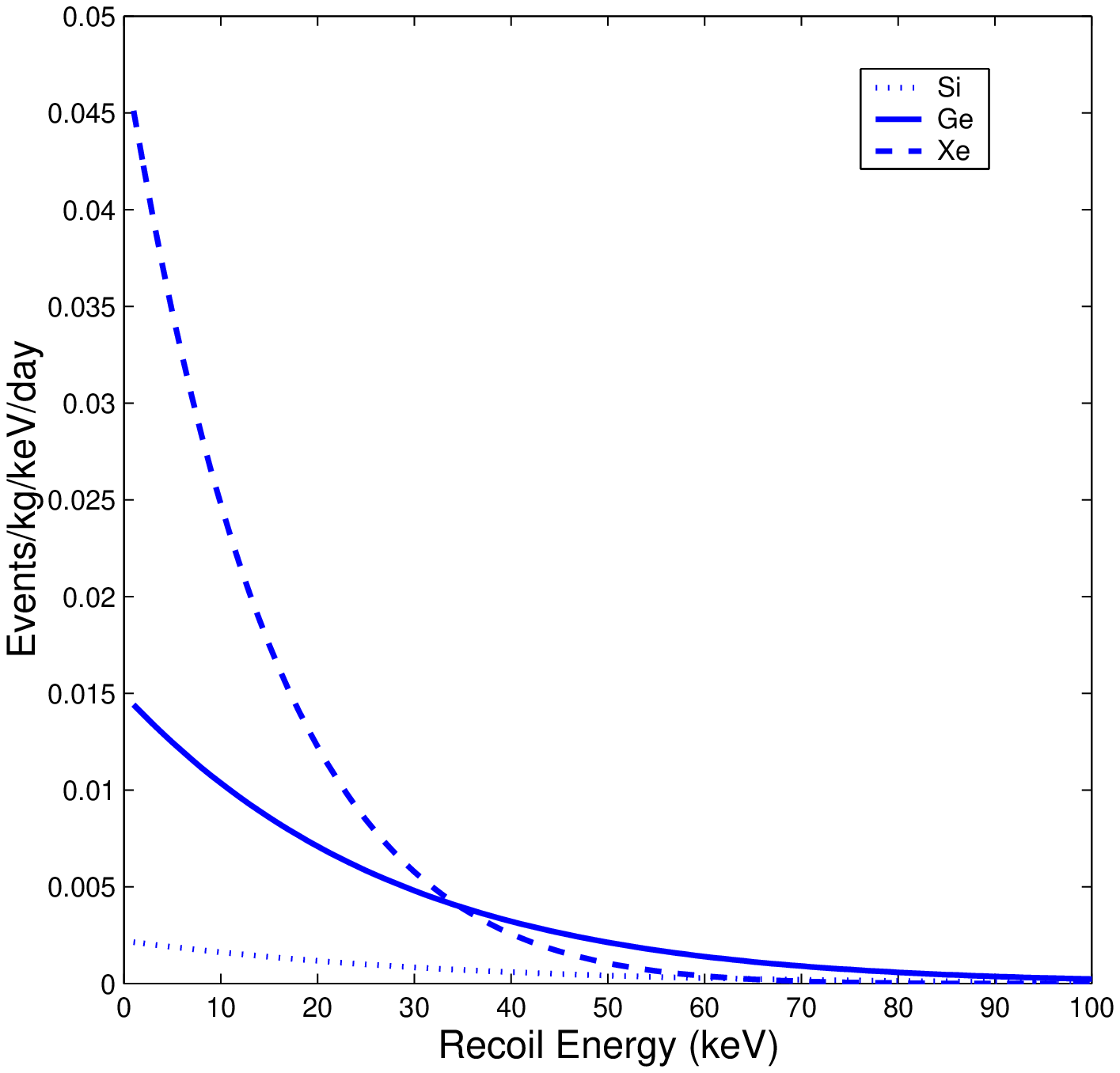}
    }
  }
 \caption{{\it
      On the left}: Nuclear form factors for Xe (dashed), Si (dots), Ge (solid). {\it On the right}: Differential recoil-energy spectra for Si (dots), Ge (solid), Xe (dashed)\cite{vuk}.}
    \label{fig1} 
\end{figure}

 So far, the discussion has been limited to zero-momentum transfer. 
When the momentum transfer $q=(2M_TE_R)^{1/2}$ is large enough so that the wavelength $h/q$ is 
comparable to the size of the nucleus, coherence is lost and the cross section begins to 
decrease. This can usually be described by including a multiplicative form factor which depends 
on the type of WIMP-nucleon interaction (spin-dependent or spin-independent) as well 
as the nuclear structure: $\sigma(q2)=\sigma_0 F^2(q^2)$. Figure \ref{fig1} (left) shows the nuclear form 
factors calculated for various materials used in WIMP search experiments. Since different 
experiments use different target nuclei, it is preferable to report the recoil energy 
spectrum referred to nucleons (e.g. proton) rather than the nucleus, to allow easy 
comparison between experiments. If the WIMP-nucleon interaction is spin-independent, the 
contribution of various nucleons will be added coherently. Two corrections are 
important: The cross-section scales as $\mu_T=M_TM_W/(M_T+M_W)$ and the WIMP-nucleus coupling scales as $A^2$. Hence,

\begin{equation}
\sigma_{\mathrm{WIMP-Nucleus}} = \sigma_{\mathrm{WIMP-p}}\frac{\mu^{2}_{\mathrm{Nucleus}}}{\mu^{2}_{p}}A^2
\label{equ6}
\end{equation}

If the WIMP-nucleon cross-section is spin-dependent, WIMP-nucleon 
amplitudes still add coherently but the contributions of spin-paired nucleons will 
cancel each other. It is clear that the $A^2$ factor usually makes the spin-independent 
interaction dominant over the
spin-dependent. We can now summarize these descriptions in a single equation:

\begin{equation}
\frac{\partial{R} (v_e, v_{esc})}{\partial{E_R}}\bigg|_{(T,q^2)} =\frac{\partial{R} (v_e, v_{esc})}{\partial{E_R}}\bigg|_{(p,0)} 
\times F^{2}(E_R)\times S
\label{equ7}
\end{equation}

where the ($T,q^2$) subscripts denote the WIMP interaction with the 
target nucleus at non-zero momentum transfer, ($p,0$) denote the same with the proton at zero momentum
transfer, $F^2$ is the nuclear form factor, and $S$ is the scaling factor (eq. \ref{equ6}).

Figure \ref{fig1} (right) shows the expected rates on various materials 
for a 100 $\mathrm{GeV/c}^2$ WIMP with $\sigma = 10^{-42}\, \mathrm{cm}^2$. Several 
important points are apparent from these plots. First, the expected event rates are very 
low: Even for a 10 keV experimental energy threshold, one expects $<$ 0.5 event/kg/day. 
Therefore, a very important goal of the direct detection experiments is to understand 
and to suppress various types of background. Second, due to the exponential nature of 
the spectrum, the majority of the signal is at very low energies. Therefore a 
low-energy threshold is essential. Third, despite the significant $A^2$ advantage of heavy 
nuclei (e.g. Xe), the nuclear form factor (Figure \ref{fig1} left) suppression makes them less 
optimal at E$>$20 keV. Therefore, Xe-based experiments would be more advantageous only if they could
decrease the experimental threshold very low (which is not easy, as 
described later in this paper).

 Thus the main goal of a WIMP search experiment is to produce 
detectors with extremely low background and very low threshold ($<$15 keV) using materials with the best
sensitivity to WIMPs. Due to the small interaction cross-section, a 
high mass - or, more accurately, a high exposure ($M\times T$) - is desirable. This can be 
obtained using large detector masses and long exposure times. This requires a high 
stability of the readout system (in particular, readout threshold) to various environmental 
conditions. 

 Currently we can classify direct WIMPs search experiments into two different 
categories: The first category, the first developed historically, focuses on building a 
detector with high mass and to passively reduce background by shielding the detector active 
region at deep underground sites. The hope is that after long exposures the 
sensitivity to the WIMP signal rises above background and eventually one can hope to detect 
the cosmological signature (eq. \ref{equ2}). However, because the signal-to-background 
discrimination is of statistical nature, the sensitivity of this method only increases 
with $(M \times T)^{1/2}$. As representatives of this category of experiments, we discuss in more 
detail DAMA-NaI and ZEPLINI in the following sections. The second category of direct WIMP 
search experiments focuses mainly 
on an event-by-event discrimination of signal against background. As described at 
the beginning of this chapter, WIMPs interacting with the nucleus cause nuclear-recoils 
while most of the radioactive background (electromagnetic interaction) interact with 
electrons, giving electron-recoils. It has been shown (see section 2.3) that with a proper detector 
design one can distinguish the two types of events with a very high efficiency. 
Compared to the first category of experiments, the sensitivity is now enhanced in direct 
proportion to the exposure. As we will see, the experiments using this method obtained the best WIMP sensitivities.

\subsection{DAMA}
The DAMA project was begun in 1990 by an Italian group at Gran Sasso 
underground laboratory \cite{belli}. This elegant project is based on highly radiopure 
NaI(Tl) scintillator detectors shielded rigorously from radioactive background.
The scintillation properties of NaI have been studied extensively for 
nuclear physics instruments. In particular, it has been shown that nuclear-recoil 
events can produce scintillation photons. It is possible to purify the crystals to 
achieve very low levels of background. The scintillation-yields are fairly low (0.3 for Na and 
0.09 for I), leading to a recoil-energy threshold of $\sim$20 keV for I, which is the more 
interesting nucleus for the WIMP-search due to its large $A^2$ factor. The group also showed 
that there is a slight difference between the pulse shapes produced by nuclear-recoil events and those
produced by electron-recoil events. Though the latter factor could 
help to statistically discriminate WIMPs against radioactive background, it has been 
ignored in the DAMA data analysis due to low efficiency.

\begin{figure}[!h]
\begin{center}
\includegraphics[height=9cm]{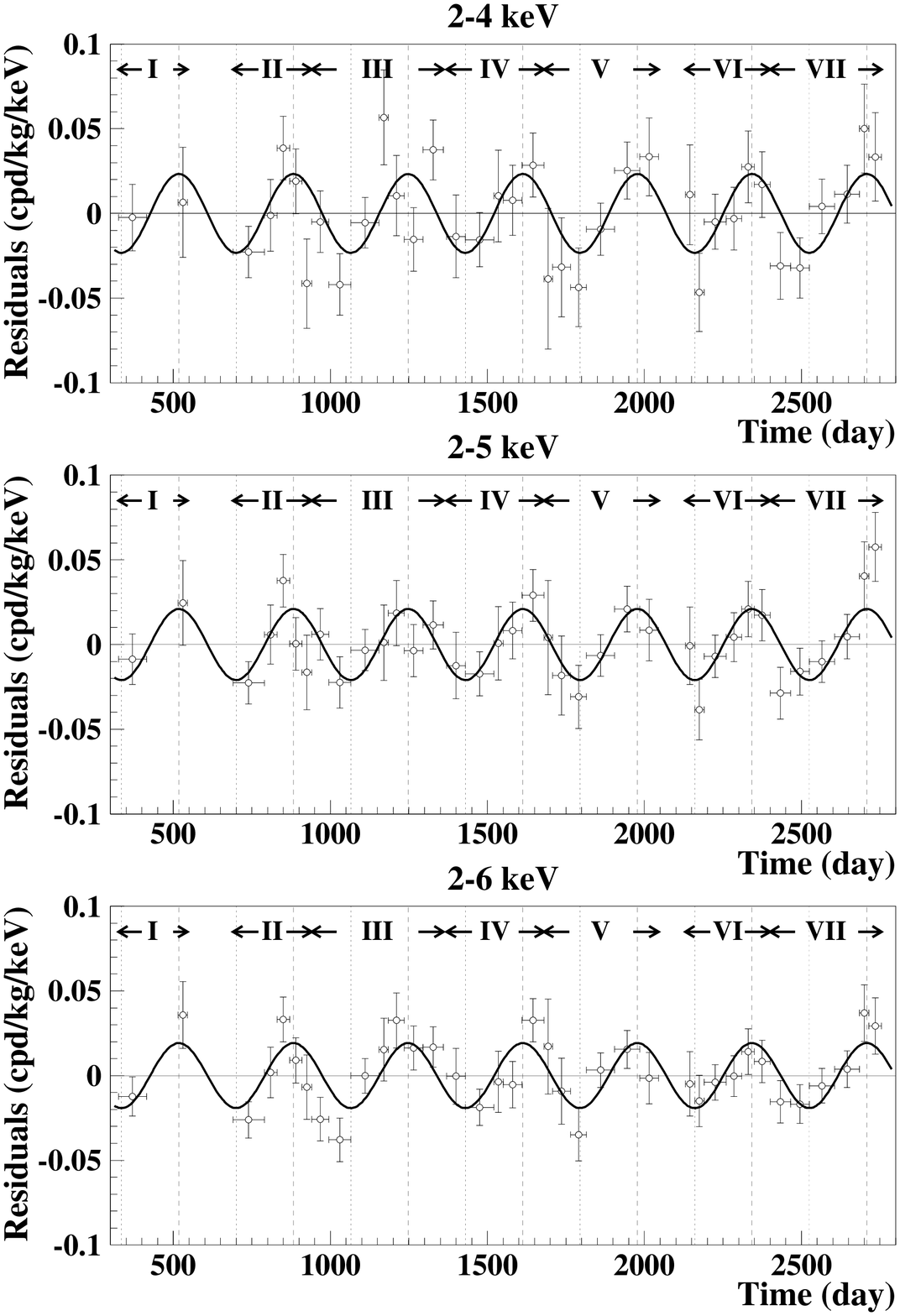}
\includegraphics[height=6cm]{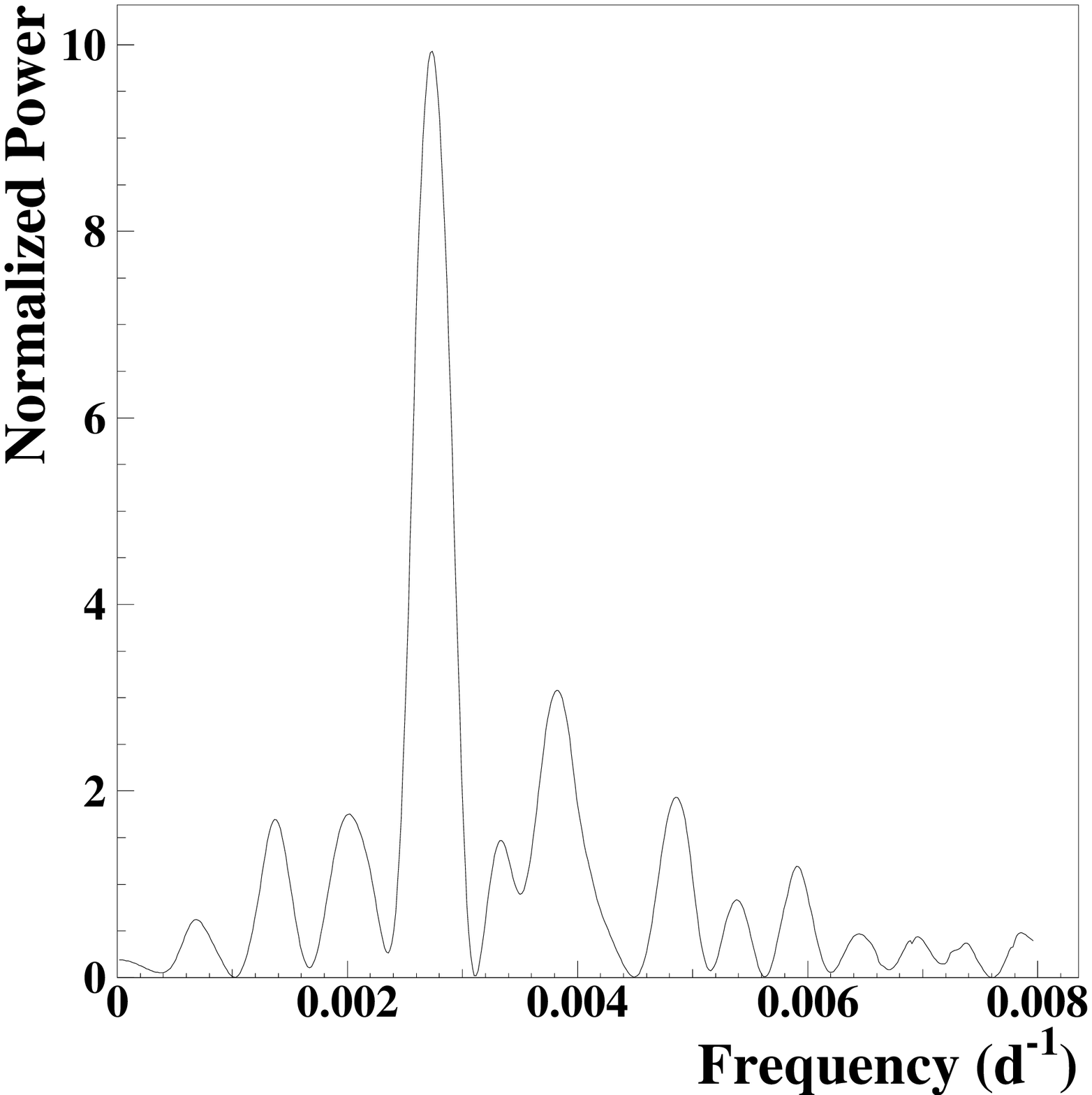}
\end{center}
\caption{{\it On the left}: The DAMA experiment's annual modulation of the residual rate (total rate minus constant) in the (2-4), (2-5), and (2-6) keV energy intervals
as a function of the time over 7 annual cycles (total exposure 
107731 kg $\times$ day); end of data taking July 2002. {\it On the right}: Power spectrum of the measured (2-6) keV modulation; the principal mode corresponds well to a 1 year period \cite{dama}.} 
\label{fig2}
\end{figure}


 The DAMA experiment is placed among the first category of dark matter 
experiments (see the end of section 2), which require a large detector 
exposure (107,731 kg-day over 7 years of operation). If WIMPs form a halo embedding our 
galaxy, as the Sun is moving through the WIMP halo, the Earth should feel a wind of WIMPs 
with strength a function of its position in its orbit. Hence, one would expect to 
observe an annual modulation of the interaction rates. In 2000, using a five-year 
exposure, the DAMA collaboration claimed to observe a 6.3 $\sigma$ C.L annual modulation in 
WIMP-proton elastic scattering. Recently, DAMA confirmed the observation \cite{dama} by 
adding the results from two more annual cycles (Figure \ref{fig2}). While the DAMA evidence for the annual
modulation is clear, its interpretation is more questionable. Most of 
the modulation signal comes from the lowest energy bins (2-6 keV), where understanding the 
efficiencies is particularly important. Although DAMA performed a study of the various possible
systematic effects, some doubts remain that the signal may be caused 
by some other, less interesting, effects. The doubts are further fueled by the fact that 
there are three experiments (namely CDMS, Edelweiss, and ZEPLIN I) which have 
explored the same parameter space and found no signal.

Several upcoming experiments may help resolve the conflict: NaIAD 
(Boulby mine, UK)\cite{ahmed} has 65 kg of NaI crystals and is already acquiring data. ANAIS 
(Canfranc, Spain) \cite{morales} plans to use a detector mass of 107 kg of NaI, and has currently 
successfully tested a prototype. Both of these experiments should test the entire signal  region claimed by 
DAMA in the coming 2-3 years.

\subsection{Xe-based experiments}

 Liquid Xe can also be used to search for a direct WIMP signal. Xe has 
an obvious advantage over other materials due to its large $A^2$ (A=131) factor. 
However, as shown in Figure \ref{fig1}, this advantage is partially offset by the form factor 
suppression. Xe-based detectors would be particularly advantageous if their 
nuclear-recoil energy threshold can be reduced below 20 keV.

 The signature of an interaction in Xe is twofold. First, there is 
electron excitation of Xe atoms, which leads to scintillation. Second, there is ionization of 
Xe atoms. The two signals can be used to discriminate against the electron-recoil 
events. In the absence of an electric field, the electron and ionized Xe recombine, producing 
secondary scintillation. The timing of the two scintillation pulses (nanosecond scale) differs 
for electron and nuclear-recoils, so pulse-shape discrimination can be used to 
suppress the electron-recoil background. This technique is used by ZEPLIN I (Boulby mine, UK) \cite{hart}. 
Their preliminary result is comparable with the Ge-based 
experiments CDMS (SUF) and Edelweiss, and incompatible with the signal region claimed by 
DAMA (Figure \ref{fig4}). However, these results are not based on an {\it in situ} neutron calibration.

 Alternatively, an electric field may be used to extract the electrons 
of the ionization signal. Such techniques are being investigated for ZEPLIN II and 
ZEPLIN III (both at the Boulby mine, UK) \cite{xe}. Possibly the most important advantage of Xe detectors is their 
scalability to large detector masses. Such experiments (ZEPLIN IV and 
XENON \cite{aprile}) are in the proposal stage. However, the dependence of the basic 
parameters, the ionization-yield and the scintillation-yield, on the energy and on the type of 
recoil are yet to be demonstrated.

\subsection{Low temperature detectors: Solutions to event-by-event discrimination}

The cryogenic calorimeters are, so far, the technologies best able to 
meet the two necessary requirements for a WIMP detector: Low threshold ($<$10keV) 
and good energy resolution ($<$100 eV). A cryogenic calorimeter consist of a dielectric 
crystal ($\mathrm{Al}_2\mathrm{O}_3$, Ge, Si, $\mathrm{CaWO}_4$, etc.) cooled to temperatures as low as 0.01 $^{\circ}\mathrm{K}$. Because the heat
capacity $C$ of a crystal varies as $(T/\Theta_D)^3$, ($\Theta_D$ is the Debye 
temperature, e.g. 374 $^{\circ}\mathrm{K}$  for Ge) at very low T a  small energy deposit from a particle interaction 
could significantly ($10^{-5}$ $^{\circ}\mathrm{K}$) change the temperature of the absorber ($\Delta T=E/C$). A properly-attached
thermometer (Mott-Anderson insulator or superconductor at its $Tc$) 
could thus measure the deposited energy. This is the best calorimetric measurement 
because all primary excitations due to the particle interaction will eventually transform 
into thermal excitations. From another point of view, at very low temperatures the 
phonon (lattice vibration quanta) content of the crystal thermal bath is very low, 
and thus the out-of-equilibrium (athermal) phonons produced after an interaction may be 
easily distinguished and counted in order to measure the deposited energy. This combined 
with the fact that the excitation energy to create phonons is very low, $\sim 10^{-5}$ eV, 
compared to $\sim$1eV for conventional semiconductor detectors, makes cryogenic detectors the 
best calorimeters at low energies yet developed.

 The above introduction suggests two distinct (but physically related) 
methods of calorimetric measurement. One method consists of measuring the 
detector temperature after it reaches the equilibrium state (at higher $T$ of course) after 
the interaction. In this case, $\Delta T =E/C$ will directly measure the energy of the interacting 
particle.  The second method, which requires a more elaborated readout system, is based on 
measuring the energy content of the athermal phonons created after an interaction, 
under the assumption that athermal phonons are proportionally produced and detected. As the athermal 
phonons carry information about the history of the event, the second method is more 
advantageous when it becomes important to reconstruct the history of an event in the 
detector.  For example, it is often possible to reconstruct the location of an event in the 
detector, which in some cases is very important as we will see later in this paper. The 
signal amplitude in the first
method depends on the mass of the detector as $C \propto M$, which limits the mass of the
detectors. The second method does not suffer from this limitation, as 
long as the lifetime of the athermal phonons  in the detector is longer than the response 
time of the readout system.

Calorimetric measurement alone is not enough to discriminate nuclear 
recoils (WIMPs) from electron-recoils (radioactive background), as the energy 
deposited does not depend on the type of interaction. However, it has been shown \cite{shutt1} that in 
semiconductor crystals (Ge, Si, etc.), the ionization-yield (charge/recoil energy) differs 
significantly between an electron-recoil and a nuclear-recoil. In Ge crystals, for example, 
the ionization-yield is 3 times bigger for an electron-recoil than for a nuclear-recoil. Figure 3 (left) shows the calibration results for one 
of the CDMS detector. Therefore, by simultaneously 
measuring ionization and phonons signal, one can obtain an event-by-event 
discrimination between a WIMP signal and the background. The CDMS and Edelweiss experiments use this
method of detection and they are currently presenting the best 
sensitivities to WIMPs. This discrimination based on the ionization-heat measurement fails 
when an event occurs very close to the detector surface, in the ``ionization dead-layer'' \cite{shutt2}.
The charge collection for such event could be incomplete, which could cause 
electron-recoil misidentification. By measuring athermal phonons, CDMS is able to 
identify and reject events occurring very close to the surface based on timing parameters 
of the phonon pulse \cite{mandic1}. The Edelweiss experiment currently uses NTD-based heat 
sensors sensitive only to overall changes in temperature, and is unable to localize 
events in this manner. However, the group's recent studies of position-sensitive 
ionization-heat detectors look very promising \cite{mirabol},\cite{marnieros}.

 Scintillation-yield (light/recoil energy) could also differ between 
electron-recoils and nuclear-recoils. The CRESST experiment is  based on simultaneous measurement of
scintillation and ionization. There are also experiments, such as CUORE and
CUORCINO \cite{cuore}, which are based on heat measurement alone. The complex techniques
involved in low temperature devices make such detectors difficult to 
scale to large masses. The main challenge of the above mentioned experiments is to 
increase the mass of the detectors without compromising their sensitivity.

\subsubsection{CDMS}

CDMS uses ZIP (Z-dependent Ionization Phonon) detector technology to 
detect WIMPs \cite{irwin}. ZIPs are disc-shaped (76 mm in diameter, 10 mm high) germanium 
or silicon crystals (absorber). One face of the disc is divided into four quadrants. Each 
quadrant is covered by a thin layer (350 nm) of lithographically-patterned aluminum fins 
(athermal phonon collectors) and 1024 tungsten transition-edge sensors 
($1 \mu m \times 250\mu m \times 35nm$ TES) which are evenly distributed over the surface. The Al layer
is directly sputtered on the surface of the absorber ($>80\%$ surface 
coverage) and provides a good phonon-phonon coupling between the two materials. The athermal 
phonons can pass the interface between the absorber and Al fins and directly 
relax their energy into the Al by breaking the Cooper pairs (creating quasi-particles) in the  Al. These 
quasi-particles will tunnel into and become trapped in the TES's 
(tungsten $Tc \sim 0.07$ $^{\circ}\mathrm{K}$). The tungsten TES's are 
biased at the superconducting transition temperature, and thus a small variation in 
the TES temperature (due to the quasiparticle trapping) will cause a significant change 
in the TES resistance ($\sim$10 m$\Omega$), which is then read out by a SQUID amplifier. The other face 
of the detector, which is also covered by aluminum, allows an 
electric field to be applied in order to collect the charge.

\begin{figure}[!h]
\begin{center}
\includegraphics[height=6.5cm]{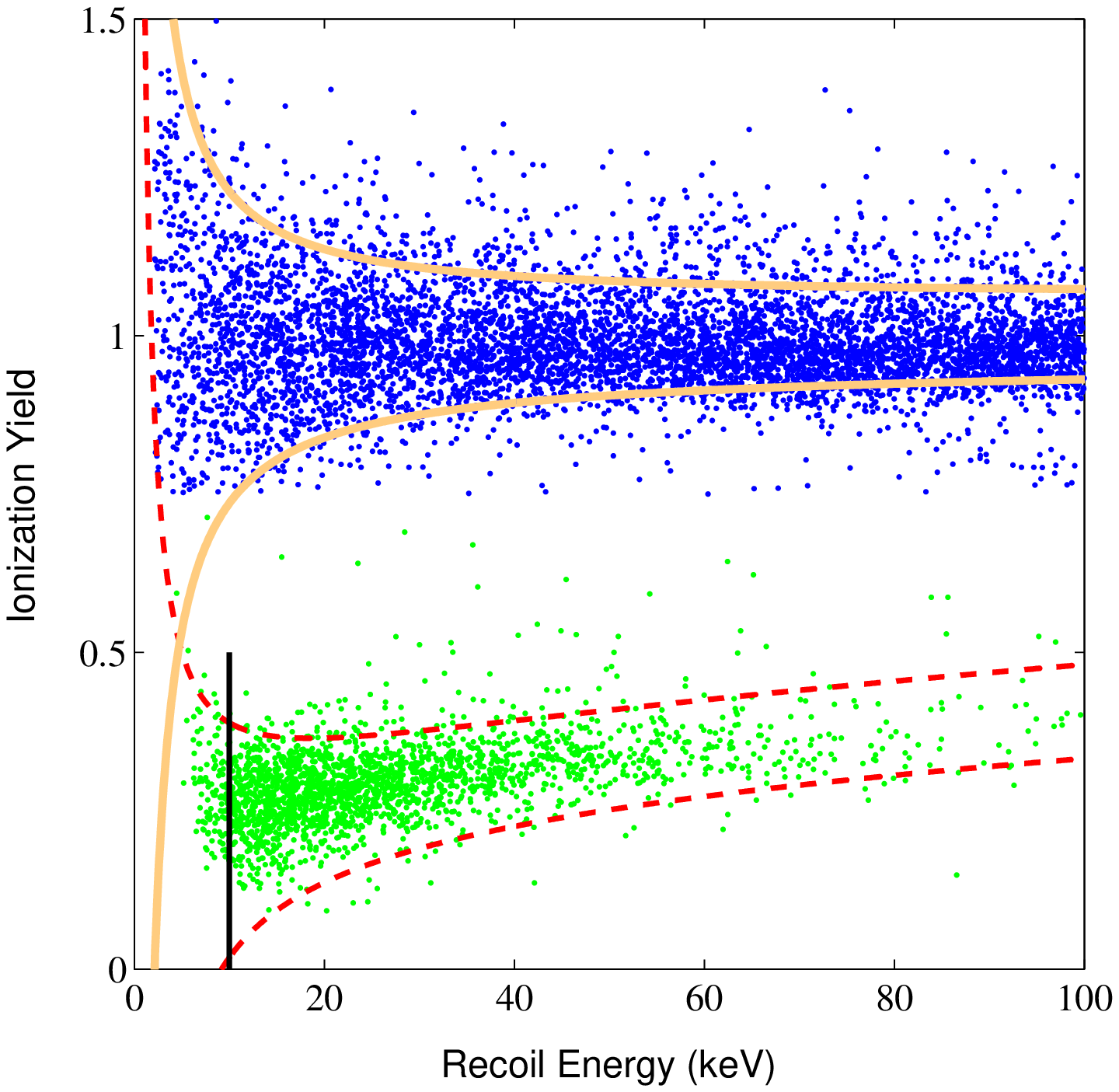}
\includegraphics[height=6.5cm]{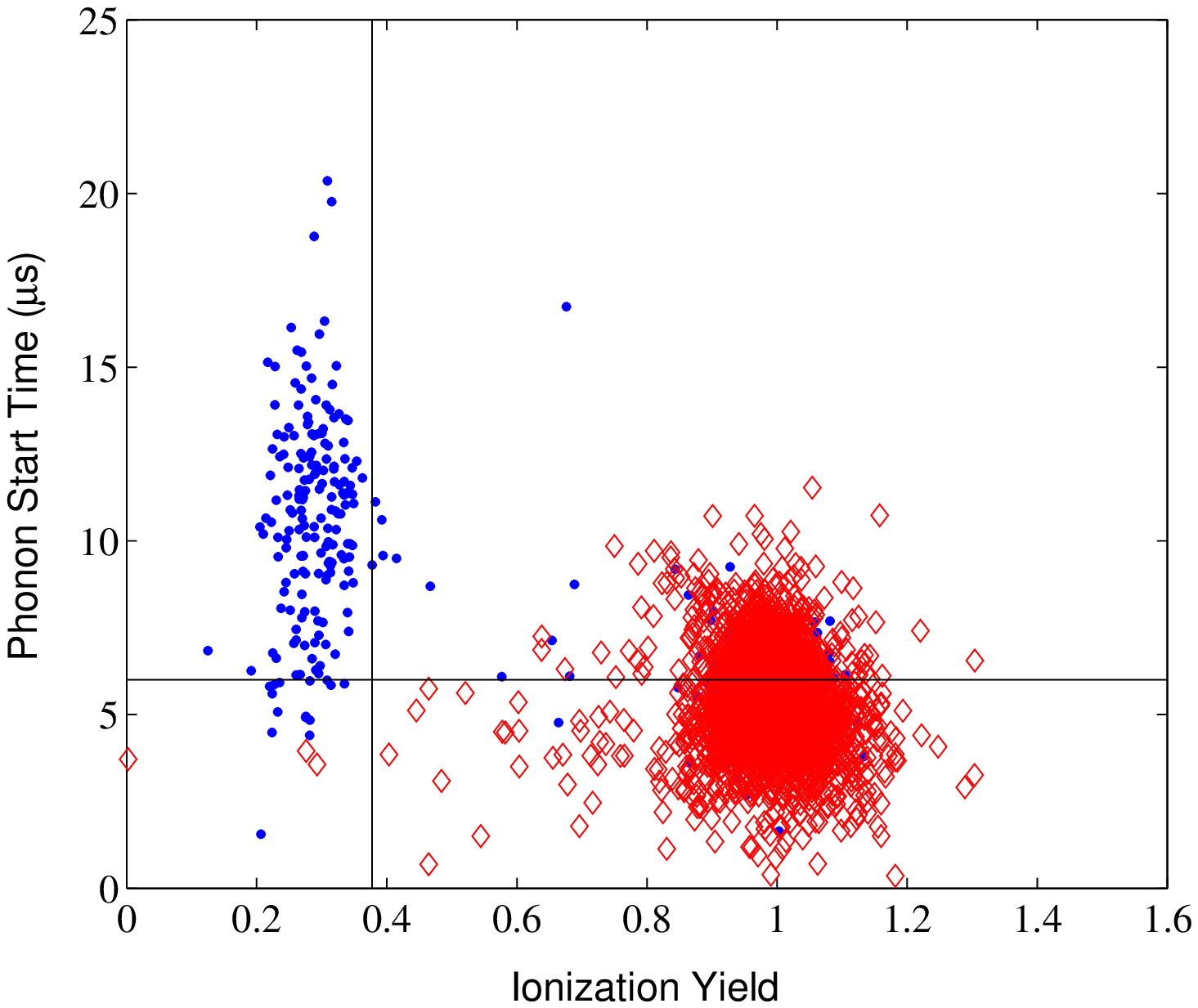}
\end{center}
\caption{{\it On the left}: Ionization-yield versus recoil energy for a typical CDMS neutron (here with ${}^{252}$Cf)  and gamma calibration. Also shown on the figure are the $\pm2\sigma$ nuclear-recoil band (dashed curve) and $\pm2\sigma$ electron-recoil band (solid curve). {\it On the right}: Phonon start time versus ionization-yield for ${}^{133}$ Ba gamma-calibration events (diamonds) and ${}^{252}$Cf neutron-calibration events (dots) in the energy range 20-40 keV in a typical CDMS detector. The diamonds that spread from yield=1 to yield=0.3 are near-surface events. Lines indicate typical timing and ionization-yield cuts, resulting in a high nuclear-recoil efficiency and a low rate of misidentified surface events\cite{cdms}.}
\label{fig3}
\end{figure}

 CDMS uses two methods to solve the near-surface event problem. First, 
an amorphous Si layer is introduced between the absorber and electrodes, which 
reduces the effect of near-surface trapping processes \cite{shutt2}. Second, the timing parameters of 
athermal-phonon signals can be used to identify the near surface events \cite{mandic1}. Figure \ref{fig3} (right) shows 
the effectiveness of using athermal-phonon signal timing parameters in rejecting the 
near-surface events. 

 Each group of six Ge (250 g) or Si (100 g) detectors is packed in a 
single ``tower'' with their corresponding cold readout electronic instruments. Five 
``towers'' are currently installed in a $\mathrm{He}_3-\mathrm{He}_4$ dilution fridge (operating $T<0.05$ $^{\circ}\mathrm{K}$) at 
Soudan underground laboratory. An overburden of 780 m of rock  reduces the 
surface muon flux by a factor of $5 \cdot 10^{-4}$. Furthermore, the detectors are shielded against 
ambient radioactivity by $\sim$0.5 cm of copper, 22.5 cm of lead, and 50 cm of polyethylene (to 
shield against neutrons). A 5-cm-thick scintillator muon veto enclosing the 
shielding identifies charged particles (and some neutral particles) that pass through it.

 Recently, CDMS published \cite{cdms} the analysis of its first Ge WIMP-search 
data (from the first ``tower'') taken at Soudan during the period October 11, 2003 
through January 11, 2004. After excluding time for calibrations, cryogen transfers, 
maintenance, and periods of increased noise, they obtained 52.6 live days with the four Ge and 
two Si detectors of ``Tower 1''. This analysis revealed no nuclear-recoil events in 52.6 
kg-d raw exposure in the Ge detectors. The data was used to set an upper limit on the 
WIMP-nucleon cross-section of $4 \cdot 10^{-43} \, \mathrm{cm}^2$   at the 90$\%$ C.L. at a WIMP mass of 60 $\mathrm{GeV}/\mathrm{c}^2$ for 
coherent scalar interactions and a standard WIMP halo (Figure \ref{fig4}). CDMS, which currently gives 
the best sensitivity to WIMPs yet attained, is now operating 2 detector towers (Tower 
1+Tower 2) and plans to run 5 towers through the year 2005. The expected sensitivity reach 
for $\sigma_{\mathrm{WIMP-nucleon}}$ is $\sim 3 \cdot 10^{-44}\, \mathrm{cm}^2$ based on 1200 kg-d  projected esposure. Also a 99$\%$-C.L. 
detection possibility is considered if  $\sigma_{\mathrm{WIMP-nucleon}} \sim 6 \cdot 10^{-44}\, \mathrm{cm}^2$.

\subsubsection{Edelweiss}

The Edelweiss experiment \cite{benoit1} is located at the LSM (French acronym for Modane
Underground Laboratory). About 1700 m of rock protect  the experiment from
radioactive backgrounds generated by cosmic rays. In the laboratory, 
the muon flux is reduced by a factor $2\cdot 10^{-6}$ compared to the flux at sea level. The 
experiment is surrounded by passive shielding made of paraffin (30 cm), lead (15 cm), and 
copper (10 cm). Edelweiss uses the same principle as CDMS for WIMP detection: Ionization-heat
discrimination. Unlike CDMS's athermal phonon sensors, the tiny rise 
in temperature due to a particle event is measured by an NTD (Neutron Transmutation 
Doped) heat sensor glued onto one of the charge-collection electrodes.

\begin{figure}[htbp]
  \centerline{\hbox{ \hspace{0.2cm}
      \includegraphics[width=6.5cm]{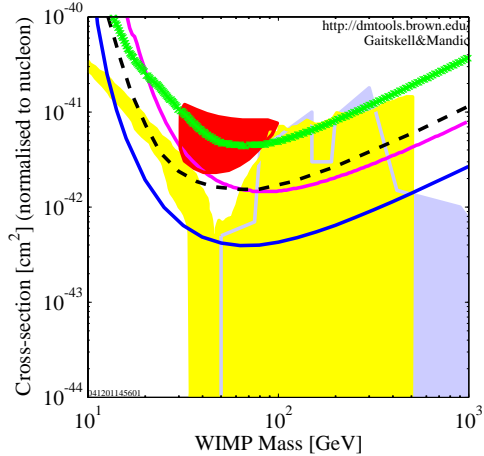}
    }
  }
 \caption{Limits on the WIMP-nucleon scalar cross section from CDMS\cite{cdms} (solid blue), Edelweiss \cite{benoit1} (solid magenta), Zeplin I preliminary \cite{zeplin}(dashed black), and CRESST (green crosses). Parameter above the curve is excluded at 90$\%$ C.L. These limits constrain several supersymmetry models, for example \cite{bottino} (yellow) and \cite{baltz} (blue). The DAMA 3 $\sigma$ signal region \cite{dama} is shown as a closed contour.}     
      
    \label{fig4} 
\end{figure}

 In 2000 and 2002, 11.6 kg-day were recorded with two different 
detectors \cite{benoit2}. In 2003, three new detectors were placed in the cryostat and 20 kg-day were 
added to the previous published data. Three events compatible with nuclear-recoils have 
been observed. However, the recoil energy of one of the events is incompatible with 
a WIMP mass $<1\, \mathrm{TeV/c}^2$. The two other events have been used to set the upper limit 
for WIMP-nucleon spin-independent interaction shown in Figure \ref{fig4}. The new limit is 
identical to the previous (11.7 kg-day), since the experiment is currently background-limited. 
The lack of an active surface-event rejection makes the distinction between nuclear-recoils 
and near-surface background events very difficult. Edelweiss is now implementing a new 
design based on NbSi thin-film Anderson insulator thermometers. The new detectors, 
which are sensitive to athermal phonons and have already demonstrated a high surface 
event rejection efficiency, will be functional during the Edelweiss II experimental stage \cite{broniat}.

 As of March 2004, the Edelweiss I experiment has been stopped to 
allow the installation of the second-stage Edelweiss II. The aim is a factor of 100 
improvement in sensitivity. A new low-radioactivity cryostat (with a capacity of 50 liters), able 
to receive up to 120 detectors, is being tested in the CRTBT laboratory at Grenoble. The 
first runs will be performed with twenty-one 320 g Ge detectors equipped with NTD heat sensors and
seven 400 g Ge detectors with NbSi thin film. With an improved 
polyethylene and lead shielding and an outer muon veto, the expected sensitivity for $\sigma_{\mathrm{WIMP-nucleon}}$ is about $10^{-44}\, \mathrm{cm}^2$.

\subsubsection{Scintillation-heat : CRESST}

The simultaneous detection of scintillation light and phonons in 
cryogenic calorimeters using scintillating absorber crystals can give a background 
suppression similar to that provided by the simultaneous measurement of ionization and light. 
Very recently it was shown \cite{coron} that a large variety of scintillating crystals ($\mathrm{CaWO}_4$, BaF, $\mathrm{PbWO}_4$, etc.) can be used in this manner. This gives this method a big advantage in 
identification of WIMP signals. The experiments using this technique are CRESST II and Rosebud.
This technique has an important advantage over the Ge-based detectors 
in that it does not have surface-event problems. However, the technique also has some 
difficulties. First, rather than using PMTs to observe the scintillation signal (due to 
their high radioactive background), the current approach (taken by CRESST II \cite{altman}) is to use 
a second, phonon-mediated detector adjacent to the primary detector. The light 
collection is relatively poor, resulting in an energy threshold of 15-20 keV. Second, there are 
three nuclei in the crystal, all of which could potentially interact with the WIMPs. The 
scintillation-yield produced by the three nuclei has yet to be studied carefully, making 
event interpretation difficult. The goal of CRESST II is to build a 10 kg detector 
consisting of 300 g crystals  to reach a sensitivity for  $\sigma_{\mathrm{WIMP-nucleon}}$ of the order of $10^{-44}\, \mathrm{cm}^2$.

\section{Indirect WIMP Searches}

We now review the current state of the various WIMP indirect search 
methods. Indirect detection experiments search for products of WIMP annihilation in 
regions that are expected to have relatively large WIMP concentration. Examples of 
such regions are galactic centers, the center of the Sun, or the center of the Earth, 
where the WIMPs are expected to be gravitationally captured. Such searches assume that 
the WIMP is its own antiparticle (as predicted by SUSY models) or that equal numbers of 
WIMPs and anti-WIMPs are present. Higher WIMP density gives a larger annihilation 
signal, which can be manifested as a flux of $\gamma$-rays, neutrinos, or antimatter 
(positrons or anti-protons) produced in the WIMP-annihilation. We discuss these 
possibilities in some detail.

\subsection{$\gamma$-rays}

WIMP-annihilation can produce $\gamma$-rays in several different ways. 
First, a continuous spectrum is produced from the hadronization and decay of $\pi_0$'s
produced in the cascading of the annihilation products. Second, $\gamma$-ray spectral lines are 
produced by the annihilation channels in which $\gamma$'s are directly produced, such as $XX \rightarrow \gamma \gamma$ (producing 
a line at $M_X$) and $XX \rightarrow \gamma Z$ (producing a line at $M_X(1-M^{2}_{Z}/M^{2}_{X})$). Observing such lines would 
be a clear detection of WIMP annihilation. Such $\gamma$-rays could be produced close to the
galactic center. Although the production rates are relatively low, a 
large halo density may compensate sufficiently to make such signals observable.

 Ground-based experiments rely on Atmospheric Cerenkov Telescopes (ACTs), which
detect the Cerenkov light emitted by the shower produced by a $\gamma$-ray interacting at the top of the atmosphere. Some experiments, such as CELESTE (France) \cite{cenbg} and
STACEE (New Mexico) \cite{stacee}, use the large mirrored areas used by solar 
power plants. These two experiments are sensitive to 20-250 GeV $\gamma$-rays. A number of 
experiments use dedicated mirrors or arrays of mirrors with a detector in the focal point:
CANGAROO (Australia) \cite{icrhp}, VERITAS (Arizona) \cite{veritas}, CAT (France) \cite{cat}, HESS
(Namibia) \cite{hess}, HEGRA (Canary Islands, dismantled) \cite{hegra}, and MAGIC 
(Canary Islands) \cite{magic}. Such experiments are typically sensitive to 100 GeV - 10 TeV $\gamma$-rays. They are also capable of distinguishing (usually at $>$ 99$\%$  efficiency) between 
the showers caused by $\gamma$-rays and those caused by cosmic rays (their dominant 
background). Finally, there are satellite-based experiments: EGRET \cite{hartman} completed its 
mission and observed $\gamma$-rays in the 20 MeV - 30 GeV energy range, and GLAST \cite{glast} is scheduled to
launch in 2006 and observe $\gamma$-rays of energies 10 MeV - 100 GeV.

 Recently, two experiments have observed an excess flux of $\gamma$-rays
coming from the galactic center. VERITAS \cite{kosack}, operating at the Whipple 10 m 
telescope on Mt. Hopkins, Arizona, observed an integral flux of $1.6\pm0.5\pm0.3\cdot 10^{-8} \mathrm{m}^{-2}\mathrm{s}^{-1}$  with the 
energy threshold of 2.8 TeV. CANGAROO \cite{tsuchi}, with a lower threshold of 250 GeV, made a $\sim 10 \sigma$ detection
of the $\gamma$-ray source in the galactic center over the range 250 GeV -2.5 TeV.
However it is difficult to reconcile \cite{hooper} the results of CANGAROO and 
VERITAS. The spectrum measured by CANGAROO is consistent with a WIMP mass of 1-3 TeV, while
VERITAS, with its energy threshold of 2.8 TeV, requires a much heavier WIMP.
Moreover, very high annihilation rates are required for this signal 
to be explained by WIMP annihilation. This implies very high annihilation cross-section 
and very high dark matter concentration at the galactic center. We note the possibility 
that these observations could also be explained by astrophysical sources, in particular the 
black hole at the galactic center. Results from the HESS experiment are expected in the 
near future - with four telescopes - HESS is expected to be more sensitive in the 
direction of the galactic center and to have superior angular resolution.

\subsection{Neutrinos}

Although WIMPs are expected to scatter very infrequently, they do 
scatter off of nuclei in the Sun or the Earth, lose, and become gravitationally bound. Hence, 
the density of WIMPs at the center of the Earth or the Sun can be considerably 
larger than in the halo, implying higher annihilation rates. Neutrinos produced in such 
WIMP-annihilations would penetrate to the Earth's surface, or escape from the Sun. The 
neutrinos can be produced both directly $XX\rightarrow \nu \bar{\nu}$ and indirectly $XX\rightarrow f \bar{f}$, where the 
fermion $f$ can decay and emit a neutrino. Hence, the energy spectrum is expected to be 
continuous, rather than a line, but it is expected to extend up to the WIMP mass. If the 
neutrino interacts with rock sufficiently close to the Earth's surface, the products of the 
interaction may be detectable. The muon neutrinos are, hence, of the most interest, because their 
interactions produce muons which can travel considerable distance through the rock and 
reach a detector (electrons are absorbed at very short distances). The muon neutrinos 
can be detected at the surface of the Earth, usually using dedicated solar or 
atmospheric neutrino detectors. In practice, one searches for upward-going muons - for the 
high-energy neutrinos, the muons produced are well-collimated with the original neutrino 
direction and carry much of the original neutrino's energy. Hence, one can search for the 
upward-going muon signal with a high-energy-threshold detector. The only known background are
atmospheric neutrinos produced in the cosmic rays interactions with 
the atmosphere at the opposite side of the Earth.

 Experiments designed to study solar or atmospheric neutrinos can also 
be used to look for the WIMP-annihilation neutrino signal. At the moment, none of the 
experiments has observed excess neutrinos from the Earth or the Sun, but several 
experiments have determined upper bounds on their flux: Baksan (neutrino experiment in 
Caucasus, Russia) \cite{boliev}, SuperKamiokande (atmospheric neutrino experiment in Japan) \cite{desai}, MACRO
(liquid scintillator neutrino experiment in Italy) \cite{ambrosio}, and AMANDA 
II (ice Cerenkov detector at the South Pole) \cite{ahrens}. These experiments are just 
beginning to probe the theoretically-allowed regions in supersymmetric WIMP models. Future 
experiments, such as ANTARES \cite{antares} and Lake Baikal \cite{baikal}, as well as future runs of 
AMANDA II and IceCube, are expected to improve the sensitivity to WIMP-annihilation 
neutrinos by $\sim$2 orders of magnitude.

\section{Conclusion}

Direct detection experiments have already explored the regions of the 
most optimistic SUSY models. Despite their lower exposures ($\sim$50 kg-day, compared to 
110,000 kg-day), event-by-event discrimination methods are currently giving the best 
sensitivities to the WIMP-nucleon scalar scattering cross-section. Extremely high 
discrimination combined with large mass seems to be the only solution for the next generation 
of direct detection experiments. The two-order-of-magnitude increase in the sensitivity 
of next-generation experiments will explore the core of many SUSY models in the next few 
years. Indirect detection will be complementary, but hardly competitive, for low-$\sigma$ scalar WIMP
detection. When combined with accelerator (LHC) results, the next 
generation of direct detection experiments may soon let us pinpoint the nature of the dark matter.

\end{document}